\documentclass[prl,aps,showpacs,twocolumn,floatfix]{revtex4}
\usepackage{epsf}
\usepackage{amssymb,amsmath}
\usepackage[dvips]{color}
\newcommand {\be}{\begin{equation}}
\newcommand {\ee}{\end{equation}}
\newcommand {\bea}{\begin{eqnarray}}
\newcommand {\eea}{\end{eqnarray}}

\begin{document}

\title{Quantum phase transitions in the Fermi-Bose Hubbard model}
\author{L.~D. Carr and M.~J. Holland}
\affiliation{JILA, National Institute of Standards and Technology
and Department of Physics, University of Colorado, Boulder, CO
80309}
\date{\today}

\begin{abstract}
We propose a multi-band Fermi-Bose Hubbard model with on-site
fermion-boson conversion and general filling factor in three
dimensions. Such a Hamiltonian models an atomic Fermi gas trapped
in a lattice potential and subject to a Feshbach resonance. We
solve this model in the two state approximation for paired
fermions at zero temperature.  The problem then maps onto a
coupled Heisenberg spin model. In the limit of large positive and
negative detuning, the quantum phase transitions in the Bose
Hubbard and Paired-Fermi Hubbard models are correctly reproduced.
Near resonance, the Mott states are given by a superposition of
the paired-fermion and boson fields and the Mott-superfluid
borders go through an avoided crossing in the phase diagram.
\end{abstract}

\pacs{}

\maketitle

The experimental investigation of cold atomic gases is proceeding
rapidly. Both bosons and fermions have been brought to quantum
degeneracy~\cite{davis1995,jin1999}. They have been trapped in the
sinusoidal lattice potential created by an optical standing wave
of two counter-propagating lasers~\cite{anderson1998,kohl2004}. In
the tightly bound regime of this potential, the Bose-Hubbard
Hamiltonian has proven to be a useful model to describe the
transition from a superfluid, in which the atoms are delocalized,
to a Mott insulator, which has an integer number of atoms at each
lattice site~\cite{fisher1989,sachdev1999,jaksch1998,greiner2002}.
Recently, the successful implementation of Feshbach resonances in
degenerate Fermi gases has enabled the experimental study of the
Bardeen-Cooper-Schrieffer (BCS) to Bose-Einstein condensate (BEC)
crossover in the continuum, a long standing theoretical
problem~\cite{nozieres1985,timmermans2001}. Initial evidence of a
new superfluid state has been found in the strongly interacting
regime~\cite{regal2004}. It is a logical next step to study such a
crossover in an atomic Fermi gas trapped in a lattice
potential~\cite{kohl2004}.

\begin{figure}[tb]
\begin{center}
\epsfxsize=7.8cm \leavevmode \epsfbox{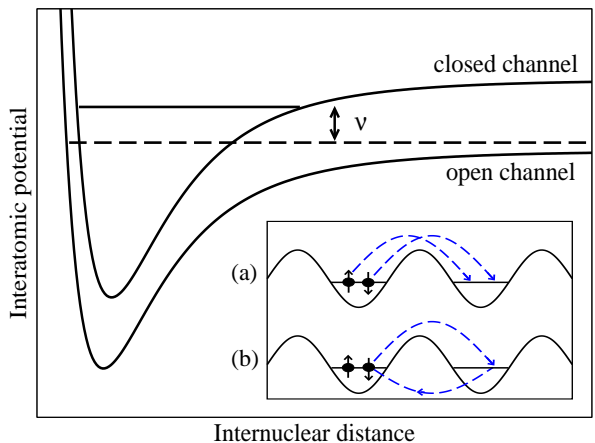} \caption{
Outer figure: pairs of fermionic atoms in the open channel are
coherently transferred into a closed channel, bosonic state via a
Feshbach resonance. Inset: second order degenerate perturbation
theory in the limit $J_f\ll V_f$ leads to two hopping events on
the lattice, (a) pair hopping, and (b) a single fermion hopping to
an adjacent site and back. \label{fig:1}}
\end{center}
\end{figure}

In this Letter, we investigate the BCS-BEC crossover in the
context of the Fermi-Bose Hubbard Hamiltonian (FBHH), motivated by
this vigorous experimental activity. Hubbard models have proven
useful in experiments on BEC's~\cite{jaksch1998,greiner2002} and
are expected to be equally relevant for
fermions~\cite{hofstetter2002,kohl2004}. A phenomenological
fermion-boson conversion term in a simplified FBHH was first
suggested in the context of high temperature
superconductors~\cite{ranninger1985}, while FBHH's without
conversion have been treated in the context of cold quantum
gases~\cite{lewenstein2004}. Unlike in this earlier work, the
model we shall study includes the possibility of both classical
and quantum phase transitions~\cite{sachdev1999} in the Fermi and
Bose Hubbard limits, as well as a conversion term. In addition, we
allow the fermions to occupy multiple bands, so that the filling
factor is not constrained.  In contrast to
high-$T_c$~\cite{ranninger1985}, fermion-boson conversion is a
real physical process in cold quantum gases, where a Feshbach
resonance is used to coherently transfer fermionic atoms into a
bound two-atom bosonic state, as illustrated in Fig.~\ref{fig:1}.

The effect of the conversion term is to lock the order parameter
of the fermions and bosons together.  It thus leads to a reduction
in the number of quantum phases from four to two. The main reason
we introduce a bosonic field is to describe the the BCS-BEC
crossover: the attractive Fermi-Hubbard Hamiltonian, even in the
paired fermion limit, does not map simply onto the repulsive Bose
Hubbard Hamiltonian~\cite{emery1976}.  After proposing this new
FBHH, we solve it in detail in the limit of on-site paired
fermions~\cite{emery1976} in the two-state approximation at zero
temperature for a filling of from zero to two fermions per site.
The on-site paired-fermion limit corresponds to the experimentally
realizable case of a strongly confining potential and/or strong
interactions.  In this limit the FBHH maps isomorphically onto a
coupled Heisenberg spin model, or coupled magnets.

Consider the FBHH in the grand canonical ensemble,
\begin{align} &H=H_f+H_b+H_{fb}\label{eqn:h}\,,\\
&H_b\equiv -J_b\sum_{\langle
i,j\rangle}(b^{\dagger}_i b_j+
b_i b^{\dagger}_j)\nonumber\\
&+\frac{1}{2}V_b\sum_i n_i^b(n_i^b-1)-\mu_b\sum_i
n_i^b\,,\label{eqn:hb}\\ &H_f\equiv -J_f\sum_{\langle
i,j\rangle,s,m,m'}(f^{\dagger}_{ism}
f_{jsm'}+f_{ism} f^{\dagger}_{jsm'})\nonumber\\
&-\frac{1}{2}V_f\sum_{i,m,m'} n^f_{i\uparrow m}n^f_{i\downarrow
m'} -\sum_{i,s,m}(\mu_f-E_m)n^f_{ism}\, ,\label{eqn:hf}\\
&H_{fb}\equiv g\sum_i (b^{\dagger}_i
f_{i\uparrow}f_{i\downarrow}+b_i
f^{\dagger}_{i\downarrow}f^{\dagger}_{i\uparrow}) +
\frac{V_{fb}}{2}\sum_{i,s,m}n^b_i n^f_{ism}\,.\label{eqn:hfb}
\end{align}
Equations~(\ref{eqn:hb})-(\ref{eqn:hf}) are the usual repulsive
Bose-Hubbard and multi-band attractive Fermi-Hubbard Hamiltonians
for a uniform lattice and Eq.~(\ref{eqn:hfb}) is the fermion-boson
coupling. The symbol $\langle i,j \rangle$ denotes nearest
neighbors, while the indices $s\in\{\uparrow,\downarrow\}$ and $m$
denote the spin state and band number. The hopping or tunnelling
strengths $J_{f,b}$ and the on-site interaction strengths
$V_{f,b}$ are taken as real and positive definite.  The band gap
energy of the $m^{\mathrm{th}}$ band is $E_m$.
The strength $g$ of the interconversion term and $V_{fb}$ of the
density coupling may have either sign. The creation and
annihilation operators $f^{\dagger},f$ and $b^{\dagger},b$ satisfy
the usual commutation relations for fermions and bosons,
respectively. The number operators are defined as
$n^b_i=b^{\dagger}_i b_i,\,n^f_{ism}=f^{\dagger}_{ism} f_{ism}$.
In order to match the physical context of quantum degenerate gases
in chemical equilibrium, we require \be
\mu_b=2\mu_f+\hbar\nu\,,\label{eqn:chemequil}\ee where $\nu$ is
the detuning associated with a Feshbach resonance and we set
$\hbar=1$. The conserved quantity \be n\equiv 2\textstyle\sum_{i}
n^b_i+\sum_{i,s,m}n^f_{ism}\ee is the total number of fermions.
Eliminating $\mu_b$ by substituting Eq.~(\ref{eqn:chemequil}) into
Eqs.~(\ref{eqn:h})-(\ref{eqn:hfb}), one finds that $\mu_f$
multiplies $n$.  One can thus take $\mu_f$ as the chemical
potential of the coupled system, while $\nu$ determines the
relative number of bosons and fermions.

The FBHH of Eqs.~(\ref{eqn:h})-(\ref{eqn:hfb}) models a
pseudo-spin-1/2 system of fermions with $s$-wave interactions, as
in experiments~\cite{jin1999,kohl2004,regal2004}. In practice, the
index $s\in\{\uparrow,\downarrow\}$ represents two hyperfine
states in the level structure of an effectively fermionic alkali
atom, such as $^{40}$K or $^6$Li, scattering near threshold in an
open channel. The bosonic field represents a bound closed-channel
molecular state, $^6$Li$_2$ or $^{40}$K$_2$, which is coupled to
the fermionic field via a resonance with an unbound open-channel
atomic state, called a Feshbach resonance. A schematic is shown in
Fig.~\ref{fig:1}(a). Note that $V_{f,b}$ and $g$ are not functions
of $\nu$.  Methods for calculating the parameters $V_f, V_b$, etc.
in Eqs.~(\ref{eqn:hb})-(\ref{eqn:hfb}) from few-body atomic
physics have been described in detail
elsewhere~\cite{dickerscheid2004}. Another important assumption is
that the pairing of fermions into bosons occurs on-site. This is
physically reasonable for present experiments~\cite{greiner2004}.

We consider the limit in which $J_f \ll V_f$, which corresponds to
a strongly confining lattice~\cite{otherlimit}.  Since the lattice
height is proportional to the intensity of the lasers creating the
standing wave, this is straightforward to obtain.  Because the
on-site interactions are attractive and $s$-wave, and the hopping
is taken perturbatively, the fermions form spin-up/spin-down
pairs.  We also restrict them to be in the lowest band.  This is
the typical experiment case in three dimensions, where $10^5$ to
$10^6$ fermions are distributed among $100^3$ sites. Thus $m=1$
and $n\in [0,2]$, i.e., there are from zero to two fermions, or
zero to one fermi pair, per site. Then second order degenerate
perturbation theory maps $H_f$ onto a new spin-$1/2$ system (a
quantum XXZ model)~\cite{emery1976}: \bea H'_f=-J'_f\sum_{\langle
i,j\rangle}(\tau_{i}^+ \tau_{j}^- -n_{i}n_{j})-\mu'_f\sum_i
n_{i}\,,\label{eqn:hfprime}\eea where $J'_f\equiv 8J_f^2/V_f$ and
$\mu'_f\equiv 2\mu_f+V_f/2$. The operator $\tau_{i}^+
\equiv(\tau_{i}^-)^{\dagger}\equiv f^{\dagger}_{i\downarrow}
f^{\dagger}_{i\uparrow}$ is a pair creation/annihilation operator
and $n_{i}\equiv \frac{1}{2}(n^f_{i\uparrow
}+n^f_{i\downarrow}-1)$.  The perturbative treatment results in
two hopping-type events, as is sketched in the inset of
Fig.~\ref{fig:1}: the $\tau_i^+\tau_j^-$ term corresponds to pair
hopping, while the $n_i n_j$ term corresponds to a single fermion
hopping to an adjacent site and hopping back.

These operators obey the commutation relation
$[\tau^{\alpha}_{i},\tau^{\beta}_{j}]=2i\tau^{\gamma}_{i}
\epsilon^{\alpha\beta\gamma} \delta_{ij}$, where
$\alpha,\beta,\gamma\in\{x,y,z\}$, $\tau^{\pm}_{i}\equiv
\tau^x_{i}\pm i \tau^y_{i}$, and $\tau^z_{i}\equiv n_{i}$. Thus,
despite the fact that $\tau^{\pm}_{i}$ is a creation/annihilation
operator for fermion pairs, the $\tau$ operators obey the Pauli
spin commutation relations, not the bosonic commutation relations.
This is one reason why the attractive Fermi-Hubbard model does not
map simply onto the repulsive Bose-Hubbard model, even in the
limit of strong interactions.  A second reason is that in order to
achieve such a mapping, a sum over many bands is required, since
the internal energy of bosons composed of two fermions is much
greater than the band spacing. In contrast, the FBHH is
asymptotically able to represent both the attractive Fermi-Hubbard
and repulsive Bose-Hubbard models in a simple way.  It is
therefore a good candidate for the study of the BCS-BEC crossover.

In general, a paired Fermi Hubbard Hamiltonian can act on all
number states of the fermions. However, as in
Eq.~(\ref{eqn:hfprime}) we consider only $n\in[0,2]$, the Hilbert
space on which it operates is restricted to two paired-number
states. Thus $H_f'$ is equivalent to the Heisenberg spin
Hamiltonian, or a magnet, \be
H_{\mathrm{spin}}=-\textstyle\sum_{i,j}J_{ij}\vec{S}_i
\cdot\vec{S}_j -\vec{h}\cdot\sum_i \vec{S}_i\,,\ee where $\mu'_f$
plays the role of the magnetic field $h_z$. One therefore expects
paramagnetic and either ferro- or anti-ferromagnetic phases.  The
former correspond to the superfluid phase, while the latter are
Mott and charge-density wave (checkerboard) phases.  Similarly,
the restriction of the Hilbert space on which $H_b$ operates to
two number states leads to an isotropic Heisenberg spin
Hamiltonian (the quantum XX model~\cite{sachdev1999}). We
formulate the two-state approximation for the coupled model as
superposition states of the form
$|\psi\rangle=\prod_j|\psi\rangle_j$, where
\begin{align} &|\psi\rangle_j \equiv
|0\rangle_{j}^{b}\otimes|0\rangle_{j}^{f}\cos\theta_j
+\sin(\theta_j)\,e^{i\phi_j}\nonumber\\
&\times(|1\rangle_{j}^{b}\otimes|0\rangle_{j}^{f}\cos\chi_j+
|0\rangle_{j}^{b}\otimes|1\rangle_{j}^{f}\sin\chi_j\,
e^{i\alpha_j}) \,.\label{eqn:ansatz}\end{align} The superscripts
$b$ and $f$ refer to Fock states of bosons and fermi pairs on the
$j^{\mathrm{th}}$ site.

The two state approximation is useful in determining the
Mott-superfluid borders in the phase diagram. The Mott state is a
single number state, while the lowest order approximation of a
superfluid is a superposition of two number states.  Therefore,
Mott states occur in Eq.~(\ref{eqn:ansatz}) for
$\theta\in\{0,\pi/2,\pi\}$.  The mixing angle $\chi$ is determined
by the detuning $\nu$ in Eq.~(\ref{eqn:chemequil}). To determine
which phase is energetically favorable one evaluates
$E_{\mathrm{gs}}\equiv\langle\psi| H |\psi\rangle$. We make the
uniform approximation $\theta_{j}=\theta\,,\phi_{j}=\phi$,
$\chi_{j}=\chi\,,\alpha_{j}=\alpha$.  Then $\phi$ does not appear
in the ground state energy, while $\alpha$ can only change the
sign of $g$.  Setting $g'=\min[g\exp(i\alpha)]$, neither $\phi$
nor $\alpha$ need be considered to obtain the phase diagram.  An
important point is that the ground state is either paramagnetic
(superfluid) or ferromagnetic (Mott). It can be proven that it is
not antiferromagnetic (charge-density wave), either by setting the
angles to differ by $\pi/2$ on each site, or by making a spin
rotation in the Hamiltonian~\cite{auerbach1994}.

The Mott-superfluid borders are obtained as follows. The ground
state energy is expanded around the Mott angles
$\theta\in\{0,\pi/2,\pi\}$. The zeroth order term gives the
energy. The first order term is zero, showing that the Mott state
is always an extremum. The sign of the second order term
determines whether the Mott state is a maximum or a minimum.
Setting this equal to zero, one obtains the Mott-superfluid
borders.  One must also extremize in the mixing angle $\chi$ and
determine whether or not it is a maximum. Thus there are three
conditions: \bea\partial^2
E_{\mathrm{gs}}/\partial \theta^2&=&0\,,\label{eqn:conda}\\
\partial
E_{\mathrm{gs}}/\partial \chi &=& 0\,,\label{eqn:condb}\\
\partial^2 E_{\mathrm{gs}}/\partial \chi^2 &>& 0
\,.\label{eqn:condc}\eea Using
conditions~(\ref{eqn:conda})-(\ref{eqn:condb}) to eliminate $\chi$
and Eq.~(\ref{eqn:chemequil}) to eliminate $\mu_b$, one finds a
quartic equation in $\mu_f$.  The coefficients are functions of
$J'_f$, $J_b$, $V_f$, $V_b$, $|g'|=|g|$, and $\nu$. The solution
to this quartic equation, though lengthy, can be written in closed
analytic form. It is best understood when evaluated in limits of
the parameters and for particular values of them.

First consider the case $\nu\rightarrow \pm\infty$.  We assume a
bipartide lattice with $Z$ the number of nearest neighbors.  Then
$\chi\in \{0,\pi/2\}$ and one obtains the Mott borders  \bea
\mu_{f}/V_f &=& -1/4+(Z/2)(1-2\sigma_f)J_f'/V_f\,,
\label{eqn:mottf}\\
\mu_{b}/V_b &=& - 2 \sigma_b Z J_b/V_b\,, \label{eqn:mottb}\eea
where $\sigma_{f} \equiv \pm 1$ gives the vacuum/one-fermi-pair
and $\sigma_{b} \equiv \pm 1$  the vacuum/one-boson Mott states.
Equations~(\ref{eqn:mottf})-(\ref{eqn:mottb}) correspond to the
solutions one finds for $g=0$ in the two-state approximation. For
$\nu\rightarrow +\infty$, condition (c) shows that
Eq.~(\ref{eqn:mottf}) is a minimum and Eq.~(\ref{eqn:mottb}) is a
maximum.  For $\nu\rightarrow -\infty$, the inverse is the case.
Thus the Bose Hubbard and paired-Fermi Hubbard limits are obtained
naturally from the ansatz of Eq.~(\ref{eqn:ansatz}) in the limits
of large negative and positive detuning. The FBHH we have proposed
therefore correctly obtains the endpoints of the BCS-BEC crossover
on a lattice.

\begin{figure}[tb]
\begin{center}
\epsfxsize=7.8cm \leavevmode \epsfbox{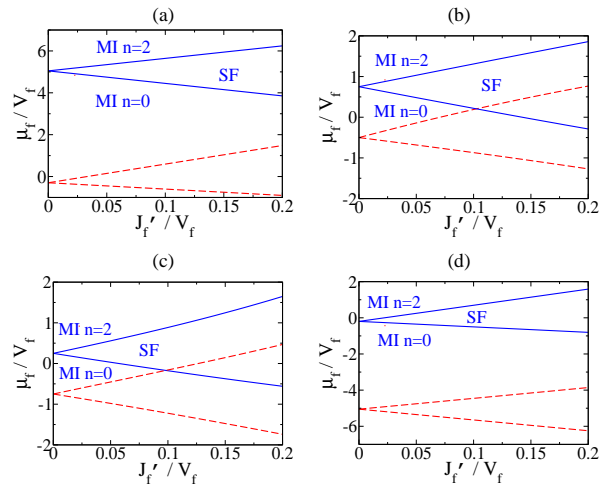} \caption{
(color online) Shown is the phase diagram for detunings (a)
$\nu/V_f=-10$, (b) $\nu/V_f=-1$, (c) $\nu/V_f=1/2$, (d)
$\nu/V_f=10$.  The blue solid curves show the Mott-superfluid
borders, while the red dashed curves show alternate extrema which
are maxima (see Fig.~\ref{fig:3}).
$\mathrm{SF}\equiv\mathrm{superfluid}$,
$\mathrm{MI}\equiv\mathrm{Mott\,\, insulator}$, $n\equiv$ fermion
filling factor.\label{fig:2}}
\end{center}
\end{figure}

\begin{figure}[tb]
\begin{center}
\epsfxsize=7.8cm \leavevmode \epsfbox{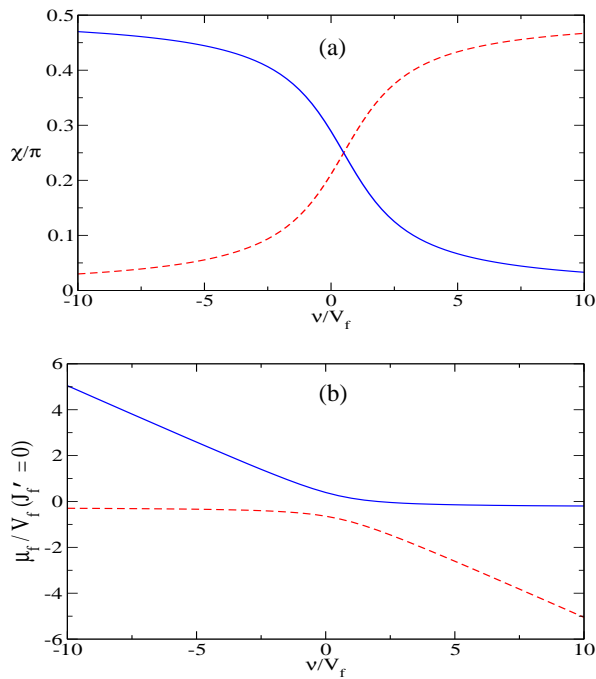} \caption{
(color online) (a) The mixing angle $\chi$ as a function of the
detuning $\nu/V_f$. (b) The $y$-intercepts in the phase diagram of
Fig.~\ref{fig:2} go through an avoided crossing as a function of
the detuning.  Blue solid curves: energy minima; red dashed
curves: energy maxima.\label{fig:3}}
\end{center}
\end{figure}

Next consider the case of the physically reasonable parameter set
$V_b=V_f$, $J_b=J'_f$, with the scaling chosen such that $V_f=1$.
The quartic equation has four roots. Two are complex and therefore
physically extraneous. The other two represent an energy minimum
and an energy maximum. There is no saddle point. The phase diagram
is shown in Fig.~\ref{fig:2} for $\nu=-10,-1,1/2,10$ and $g=1$.
The results are qualitatively the same for all $g\neq 0$. The
point $\nu=1/2$ is the actual crossover in our model, i.e., the
point at which the Mott borders become degenerate. To illustrate
this, in Fig.~\ref{fig:3}(a) is shown the mixing angle $\chi$ as a
function of $\nu$. Note the appropriate $\nu\rightarrow\pm\infty$
limits. In Fig.~\ref{fig:3}(b) are shown the $y$-intercepts of the
Mott phases from the phase diagrams of Fig.~\ref{fig:2} as a
function of $\nu$.  These go through an avoided crossing at
$\nu=1/2$. Smaller values of $|g|$ cause the avoided crossing to
become narrower. Similarly, the width of $\chi(\nu)$ in
Fig.~\ref{fig:3}(a) is proportional to $|g|$.

In conclusion, we have proposed a general Fermi-Bose Hubbard model
which describes the BCS-BEC crossover on a lattice. Our
restriction of the Hilbert space to the lowest band and paired
fermions corresponds to the experimentally realizable case of from
zero to two fermions per site in three dimensions and a strongly
confining lattice. We used a superposition ansatz
(Eq.~(\ref{eqn:ansatz})) which is relevant to both broad and
narrow Feshbach resonances, i.e., for general coupling $g$. We
found that the Paired-Fermi Hubbard and Bose Hubbard phase
diagrams appear naturally and asymptotically for large positive
and negative detuning.  We also showed that the Mott phases of the
dressed fermion and boson fields go through an avoided crossing as
the system approaches resonance.

We thank Matthew Fisher, Markus Greiner, Walter Hofstetter, and
especially Daniel Sheehy for useful discussions.  We acknowledge
the support of the Department of Energy, Office of Basic Energy
Sciences via the Chemical Sciences, Geosciences and Biosciences
Division.  LDC is grateful to the KITP for hosting him and thanks
the NSF for partial support under grants PHY99-0794 and MPS-DRF
0104447.


\begin{thebibliography}{10}
\bibitem{otherlimit}  In the opposite limit, $J_f \gg V_f$, the
fermionic field is always superfluid.  The coupling introduces
intersite correlations into the bosonic field, and thus destroys
the boson Mott state.
\end{thebibliography}
\end{document}